\documentclass[aps,prd,twocolumn,superscriptaddress,showpacs]{revtex4}
\usepackage{graphicx}	% Include figure files
\usepackage{dcolumn}	% Align table columns on decimal point
\usepackage{bm}		% bold math

\def \pisma#1#2#3#4{Pis'ma Zh.\ Eksp.\ Teor.\ Fiz.\ {\bf#1}, #2 (#3) [JETP
Lett.\ {\bf#1}, #4 (#3)]}

\def \prd#1#2#3{Phys.\ Rev.\ D {\bf#1}, #2 (#3)}

\begin{document}

%\linenumbers

\title{Measurement of the Ratio of Branching Fractions 
$ {\cal B}(D^0 \to K^+ \pi^-)/{\cal B}(D^0 \to K^- \pi^+)$
using the CDF~II Detector}

%\author{The CDF Collaboration}
\affiliation{Institute of Physics, Academia Sinica, Taipei, Taiwan 11529, Republic of China} 
\affiliation{Argonne National Laboratory, Argonne, Illinois 60439} 
\affiliation{Institut de Fisica d'Altes Energies, Universitat Autonoma de Barcelona, E-08193, Bellaterra (Barcelona), Spain} 
\affiliation{Baylor University, Waco, Texas  76798} 
\affiliation{Istituto Nazionale di Fisica Nucleare, University of Bologna, I-40127 Bologna, Italy} 
\affiliation{Brandeis University, Waltham, Massachusetts 02254} 
\affiliation{University of California, Davis, Davis, California  95616} 
\affiliation{University of California, Los Angeles, Los Angeles, California  90024} 
\affiliation{University of California, San Diego, La Jolla, California  92093} 
\affiliation{University of California, Santa Barbara, Santa Barbara, California 93106} 
\affiliation{Instituto de Fisica de Cantabria, CSIC-University of Cantabria, 39005 Santander, Spain} 
\affiliation{Carnegie Mellon University, Pittsburgh, PA  15213} 
\affiliation{Enrico Fermi Institute, University of Chicago, Chicago, Illinois 60637} 
\affiliation{Joint Institute for Nuclear Research, RU-141980 Dubna, Russia} 
\affiliation{Duke University, Durham, North Carolina  27708} 
\affiliation{Fermi National Accelerator Laboratory, Batavia, Illinois 60510} 
\affiliation{University of Florida, Gainesville, Florida  32611} 
\affiliation{Laboratori Nazionali di Frascati, Istituto Nazionale di Fisica Nucleare, I-00044 Frascati, Italy} 
\affiliation{University of Geneva, CH-1211 Geneva 4, Switzerland} 
\affiliation{Glasgow University, Glasgow G12 8QQ, United Kingdom} 
\affiliation{Harvard University, Cambridge, Massachusetts 02138} 
\affiliation{Division of High Energy Physics, Department of Physics, University of Helsinki and Helsinki Institute of Physics, FIN-00014, Helsinki, Finland} 
\affiliation{University of Illinois, Urbana, Illinois 61801} 
\affiliation{The Johns Hopkins University, Baltimore, Maryland 21218} 
\affiliation{Institut f\"{u}r Experimentelle Kernphysik, Universit\"{a}t Karlsruhe, 76128 Karlsruhe, Germany} 
\affiliation{High Energy Accelerator Research Organization (KEK), Tsukuba, Ibaraki 305, Japan} 
\affiliation{Center for High Energy Physics: Kyungpook National University, Taegu 702-701; Seoul National University, Seoul 151-742; and SungKyunKwan University, Suwon 440-746; Korea} 
\affiliation{Ernest Orlando Lawrence Berkeley National Laboratory, Berkeley, California 94720} 
\affiliation{University of Liverpool, Liverpool L69 7ZE, United Kingdom} 
\affiliation{University College London, London WC1E 6BT, United Kingdom} 
\affiliation{Centro de Investigaciones Energeticas Medioambientales y Tecnologicas, E-28040 Madrid, Spain} 
\affiliation{Massachusetts Institute of Technology, Cambridge, Massachusetts  02139} 
\affiliation{Institute of Particle Physics: McGill University, Montr\'{e}al, Canada H3A~2T8; and University of Toronto, Toronto, Canada M5S~1A7} 
\affiliation{University of Michigan, Ann Arbor, Michigan 48109} 
\affiliation{Michigan State University, East Lansing, Michigan  48824} 
\affiliation{Institution for Theoretical and Experimental Physics, ITEP, Moscow 117259, Russia} 
\affiliation{University of New Mexico, Albuquerque, New Mexico 87131} 
\affiliation{Northwestern University, Evanston, Illinois  60208} 
\affiliation{The Ohio State University, Columbus, Ohio  43210} 
\affiliation{Okayama University, Okayama 700-8530, Japan} 
\affiliation{Osaka City University, Osaka 588, Japan} 
\affiliation{University of Oxford, Oxford OX1 3RH, United Kingdom} 
\affiliation{University of Padova, Istituto Nazionale di Fisica Nucleare, Sezione di Padova-Trento, I-35131 Padova, Italy} 
\affiliation{LPNHE-Universite Pierre et Marie Curie-Paris 6, UMR7585, Paris F-75005 France; IN2P3-CNRS} 
\affiliation{University of Pennsylvania, Philadelphia, Pennsylvania 19104} 
\affiliation{Istituto Nazionale di Fisica Nucleare Pisa, Universities of Pisa, Siena and Scuola Normale Superiore, I-56127 Pisa, Italy} 
\affiliation{University of Pittsburgh, Pittsburgh, Pennsylvania 15260} 
\affiliation{Purdue University, West Lafayette, Indiana 47907} 
\affiliation{University of Rochester, Rochester, New York 14627} 
\affiliation{The Rockefeller University, New York, New York 10021} 
\affiliation{Istituto Nazionale di Fisica Nucleare, Sezione di Roma 1, University of Rome ``La Sapienza," I-00185 Roma, Italy} 
\affiliation{Rutgers University, Piscataway, New Jersey 08855} 
\affiliation{Texas A\&M University, College Station, Texas 77843} 
\affiliation{Istituto Nazionale di Fisica Nucleare, University of Trieste/\ Udine, Italy} 
\affiliation{University of Tsukuba, Tsukuba, Ibaraki 305, Japan} 
\affiliation{Tufts University, Medford, Massachusetts 02155} 
\affiliation{Waseda University, Tokyo 169, Japan} 
\affiliation{Wayne State University, Detroit, Michigan  48201} 
\affiliation{University of Wisconsin, Madison, Wisconsin 53706} 
\affiliation{Yale University, New Haven, Connecticut 06520} 
\author{A.~Abulencia}
\affiliation{University of Illinois, Urbana, Illinois 61801}
\author{D.~Acosta}
\affiliation{University of Florida, Gainesville, Florida  32611}
\author{J.~Adelman}
\affiliation{Enrico Fermi Institute, University of Chicago, Chicago, Illinois 60637}
\author{T.~Affolder}
\affiliation{University of California, Santa Barbara, Santa Barbara, California 93106}
\author{T.~Akimoto}
\affiliation{University of Tsukuba, Tsukuba, Ibaraki 305, Japan}
\author{M.G.~Albrow}
\affiliation{Fermi National Accelerator Laboratory, Batavia, Illinois 60510}
\author{D.~Ambrose}
\affiliation{Fermi National Accelerator Laboratory, Batavia, Illinois 60510}
\author{S.~Amerio}
\affiliation{University of Padova, Istituto Nazionale di Fisica Nucleare, Sezione di Padova-Trento, I-35131 Padova, Italy}
\author{D.~Amidei}
\affiliation{University of Michigan, Ann Arbor, Michigan 48109}
\author{A.~Anastassov}
\affiliation{Rutgers University, Piscataway, New Jersey 08855}
% ...
\collaboration{CDF Collaboration}
\noaffiliation

%comment out when using line numbers:
\noaffiliation

\date{\today}

%\maketitle

\begin{abstract}

We present a measurement of $R_{\cal B}$, the ratio of the branching
fraction for the rare decay $D^0 \to K^+ \pi^-$
to that for the Cabibbo-favored decay $D^0 \to K^- \pi^+$.  Charge
conjugate decays are implicitly included.  A signal of $2005
\pm 104$ events for the decay $D^0 \to K^+ \pi^-$ is obtained using
the CDF~II detector at the Fermilab Tevatron collider. The data set
corresponds to an integrated luminosity of 0.35~fb$^{-1}$ produced in
$\bar{p}p$ collisions at $\sqrt{s}=1.96$~TeV.  Assuming no mixing,
we find $R_{\cal B} = [4.05 \pm 0.21 (\mbox{stat})
\pm {}$\mbox{0.11(syst)] $\times\, 10^{-3}$}.  This measurement is consistent with
the world average, and comparable in accuracy with the best
measurements from other experiments.

\end{abstract}

% activate the following line for publication
%
% remove when using line numbers:
%
\pacs{13.25.Ft, 14.40.Lb}

\maketitle

The $D^0$ can decay to $K^+\pi^-$ either through a doubly
Cabibbo-suppressed (DCS) tree process or by oscillation (mixing) to a
$\bar{D}^0$ followed by a Cabibbo-favored (CF) tree process.  The
charge-conjugate decays, such as $\bar{D}^0 \to K^-\pi^+$, are implied
throughout this paper.  The time-dependent decay rate $r(t)$ for
$D^0 \to K^+ \pi^-$ can be written in a compact form \cite{pdg04-mix}
taking into account the experimentally established facts that the rate
for mixing is at least as small as that for the DCS decay, and that
the effect of CP violation is small.  In this formalism, and assuming
CP conservation,
\begin{equation}
r(t) \propto e^{- \Gamma t} \left[ R_D + \sqrt{R_D} y' 
(\Gamma t) + \frac{x'^2 + y'^2}{4} (\Gamma t)^2 \right].
\end{equation} 
The parameter $R_D$ is the squared modulus of the ratio of DCS to CF amplitudes.
The parameters $x'$ and $y'$ are defined in terms of the parameters $x
= \Delta m / \Gamma$ and $y = \Delta \Gamma / 2\Gamma$, where $\Delta
m$ is the difference in mass between the two mass eigenstates, $\Delta
\Gamma$ is the difference in decay width between the two mass
eigenstates, and $\Gamma$ is the average decay width.
%  The parameters $x'$ and $y'$ are defined in terms of $x$ and $y$ by
The definitions are
\begin{equation}
x' = x \cos \delta + y \sin \delta, {\rm ~and~} 
y' = -x \sin \delta + y \cos \delta, 
\end{equation}
where $\delta$ is the strong phase difference between the DCS and CF
amplitudes.  The mixing rate is small if 
$x' \ll 1$ and $y' \ll 1$.  The ratio of branching
fractions,
\begin{equation}
R_{\cal B} = {\cal B}(D^0 \to K^+ \pi^-)/{\cal B}(D^0 \to K^- \pi^+)
\end{equation}
is given by the ratio of the time-integrals of the corresponding decay rates,
\begin{equation}
\label{R-cal-B}
R_{\cal B} = R_D + \sqrt{R_D} y' + \frac{x'^2 + y'^2}{2}.
\end{equation}
Thus, if $x'$ and $y'$ are negligible, the experimentally measurable quantity
$R_{\cal B}$ can be interpreted as the theoretical parameter $R_D$.

In the limit of flavor SU(3) symmetry, $R_D = \tan^4\theta_{C}$
\cite{SU3-1, SU3-2} where $\theta_{C}$ is the Cabibbo angle, which is
measured from kaon decays. 
The world average values \cite{pdg04-mix, pdg04-vus}, $R_D = (3.62
\pm {}$\mbox{$0.29) \times\, 10^{-3}$} and
$\tan^4\theta_{C} = (2.88 \pm {}$\mbox{$0.27) \times\, 10^{-3}$},
are equal within their uncertainties, consistent with
flavor SU(3) symmetry. 
However, symmetry violation of magnitude less than the current 
measurement accuracy is possible 
\cite{Gronau-Rosner} if there are 
differences in the weak decay form factors $F^{D\pi}_0$ and
$F^{DK}_0$, or due to strong interaction resonant
intermediate states following the charm quark decay.

There is no experimental evidence either for CP violation in $D^0$
decays or for $D^0$-$\bar{D}^0$ mixing; all measurements are consistent
with \mbox{$x' = y' = 0$}.  An upper limit on the contribution to
$R_{\cal B}$ from mixing can be derived from measured upper limits
for $x' , y' \sim$ \mbox{$3 \times 10^{-2}$}~\cite{pdg04-mix} and the world average
measurement of $R_D$. A simple estimate, by substituting these values
into Eq.~(\ref{R-cal-B}), gives a contribution to $R_{\cal B}$ less
than \mbox{$2.7 \times 10^{-3}$}, a limit which is comparable to the value of
$R_D$.

In the standard model, theoretical predictions due to short-distance
weak processes are
$x', y' \sim$ \mbox{$6 \times 10^{-7}$}~\cite{Golowich:2005pt}.
However, strong
interaction effects could result in larger values, of order
$\sin^2\theta_{C}$ (=0.048) times an unknown factor which
describes the size of flavor SU(3) symmetry violation.  Thus, 
accurate measurement of $R_D$ and $\theta_C$ can establish the size
of the symmetry violation factor and make possible the prediction of
the standard model contribution to mixing.
If the standard model
contribution is small, then $D^0$-$\bar{D}^0$ mixing measurements will
be sensitive to new physics. Theories involving weak-scale
supersymmetry or new strong dynamics at the TeV scale can accommodate
large values of $x'$ and $y'$, up to the experimental limits
\cite{Burdman-Shipsey-2003}, leaving open the possibility for indirect
observation of new physics.

Until now, the most precise measurements of $R_D$ were from the
$B$ factories.  The BABAR collaboration \cite{aubert:babar} reported
$R_D$ = $[3.59 \pm 0.20 (\mbox{stat}) \pm {}$\mbox{0.27(syst)]$ \times\, 10^{-3}$}
with a DCS signal of 430 events, and the Belle collaboration
\cite{abe:belle} reported $R_D$ = $[3.81 \pm 0.17$(stat) 
\mbox{${}^{+0.08}_{-0.16}$(syst)]$ \times\, 10^{-3}$} with a DCS signal of
845 events. In this paper, using a DCS signal of 2005 events and assuming no mixing,
we report a measurement of $R_D$ with comparable precision to those of BABAR and
Belle. As in those experiments, we reconstruct the decay chain $D^{*+}
\to \pi^+ D^0$, $D^0 \to K^+ \pi^-$, where the charge of the $\pi^+$
from $D^{*+}$ decay distinguishes the $D^0$ from its anti-particle,
$\bar{D}^0$.

Our measurement uses data collected by the CDF~II detector at the
Fermilab Tevatron collider, from October 2002 to August 2004.  The
data corresponds to an integrated luminosity of 0.35~fb$^{-1}$
produced in $\bar{p}p$ collisions at $\sqrt{s}=1.96$~TeV.  CDF~II is a
multi-purpose detector with a magnetic spectrometer surrounded by a
calorimeter and a muon detector.  The CDF~II components pertinent to
this analysis are described briefly below.  A more detailed
description is found in
\cite{CDF} and references therein.  A cylindrical silicon microstrip 
vertex detector (SVX~II) \cite{cdf:silicon} and a cylindrical drift chamber 
(COT) \cite{cdf:cot}, immersed in a 1.4~T axial
magnetic field, allow reconstruction of tracks (trajectories of
charged particles) in the
pseudo-rapidity range $\left| \eta \right|~\leq~1.3$, where $\eta =
\tanh^{-1}(\cos\theta)$ and $\theta$ is the angle measured from the
beamline. The ionization signals from the COT provide a
measurement of the specific energy loss for a charged
particle, which is used for particle identification.

Events were selected in real time using a three-level trigger system
with requirements developed for a broad class of heavy flavor decays.
At level 1, tracks are reconstructed in the COT in the plane
transverse to the beamline by a hardware processor (XFT)
\cite{cdf:xft}.  Two oppositely charged tracks are required, each with
transverse momentum greater than 2~GeV/$c$.  In addition, the scalar
sum of the two transverse momenta must be greater than 5.5~GeV/$c$. 
The opening angle
$\Delta \phi$ between the two tracks in the transverse plane must be
less than $135^{\circ}$. At level 2, the silicon vertex tracker (SVT)
\cite{cdf:svt} attaches SVX II hits to each of the two XFT tracks to
increase the measurement accuracy. The transverse
impact parameter $d_0$ is defined as the distance of closest approach,
in the transverse plane, of a track to the beamline.  Each of the two
tracks is required to satisfy $120~\mu$m $\leq d_0~\leq~1.0$~mm. The
opening angle cut is tightened (compared to level 1) to 
$2^\circ \leq \Delta\phi \leq 90^\circ$.
The track pair forms a
long-lived particle candidate which is required to have a
decay length $L_{xy} > 200~\mu$m, where
$L_{xy}$ is the transverse distance from the beam line to
the candidate's vertex, projected along the total
transverse momentum of the candidate. At level 3, a conventional 
computer processor confirms the selection with a full event reconstruction.

The analysis method for determining the ratio of branching fractions
requires reconstruction of the decay chains $D^{*+} \to \pi^+ D^0$,
$D^0 \to K^- \pi^+$ (CF), and $D^{*+} \to \pi^+ D^0$, $D^0 \to K^+
\pi^-$ (DCS).
The $D^0$ candidate reconstruction starts with a pair of oppositely
charged tracks that satisfy the trigger requirements.  The tracks are
considered with both $K^-\pi^+$ and $\pi^-K^+$ interpretations.  A
third track, which is required to have $p_T \geq 0.3$~GeV/$c$, is used
to form a $D^*$ candidate when combined as a pion with the $D^0$
candidate.  The charge of this ``tagging pion'' determines whether the
$D^0$ candidate decay is CF or DCS.

To reduce systematic uncertainty, the same set of cuts is employed for
both the CF and DCS decay modes. The offline analysis cuts were
chosen to maximize the significance of the DCS signal determined from
a study of the CF signal and the DCS background.  The optimization was
performed without using DCS candidates and before the candidates were
revealed.  The DCS signal was estimated by scaling the CF signal by
the world average for $R_D$.  The DCS background was estimated from
candidates in a control region of $D^*$ invariant mass, outside a
region containing the signal. In the optimization study, the same
algorithms for data analysis were followed as for the DCS signal
determination.

We apply two cuts to reduce the background to the DCS signal from CF
decays where the $D^0$ decay tracks are mis-identified.  Mis-identification
occurs when the kaon and pion assignments are mistakenly interchanged.  This
background is characterized by a $K\pi$ mass distribution with width
about ten times that of the signal peak.  A DCS candidate that is
consistent with being a CF decay, with $K^-\pi^+$ invariant mass
within $\pm$~20~MeV/$c^2$ of the $D^0$ mass, is excluded from the DCS
signal.  This cut
rejects 97.5\% of mis-identified decays, while retaining 78\% of the
signal.  Since the analysis procedure is the same for DCS and CF
decays, a CF candidate that is consistent with being a DCS decay is
excluded from the CF signal.

A cut based on particle identification from specific ionization in the
COT also helps to reject mis-identified decays, but with a smaller
improvement to DCS signal significance than from the cut based on
invariant mass.
A variable $Z$ is defined as the
ratio of logarithms of measured to predicted charge deposition for a
single track. The prediction is based on the ionization expected for a
particle with the measured momentum and a specific hypothesis for
mass. For a pair of particles, we define
\begin{equation}
S_{K \pi} = \left( {Z_K \over \sigma_{K}} \right)^2 + \left( {Z_{\pi} \over
\sigma_{\pi}} \right)^2
\end{equation}
where the subscripts $K$ and $\pi$ indicate the particle hypotheses
for the first and second track of the pair,
and $\sigma_K$, $\sigma_{\pi}$ are the corresponding Gaussian
resolutions on $Z$.  The particle identification for the pair is
chosen by the smaller of $S_{K \pi}$ and $S_{\pi K}$.  This selection is
correct 80.2\% of the time, as measured using the CF signal that survives
the invariant mass cut.

We apply four cuts to reduce combinatoric background from prompt
particle production or from improper combinations of tracks in events
containing heavy flavor particles.  These cuts retain most of the
signal, with a small improvement in the signal significance.  Since
the $D^0$ has a long enough mean lifetime to have an observable decay
length, the decay vertex should be displaced, on average, from the
production point. We require the transverse decay length significance
$L_{xy}/\sigma_{xy}~>~5$, where $L_{xy}$ was defined earlier and
$\sigma_{xy}$ is the uncertainty on $L_{xy}$.  The $D^*$ has a short
enough mean lifetime so that it should appear to decay at its
production point.  The tagging pion and the $D^0$ candidate must be
consistent with coming from a common point based on a $\chi^2$ measure
in the transverse plane.  For the $D^*$ vertex, we also require
$\left| L_{xy}/\sigma_{xy} \right|~<~15$.  Furthermore, the tagging
pion track must have an impact parameter $d_0~<~800~\mu$m.

\begin{figure}[!tbp]
\vskip +0.4cm
\begin{center}
\includegraphics[width=3.4in]{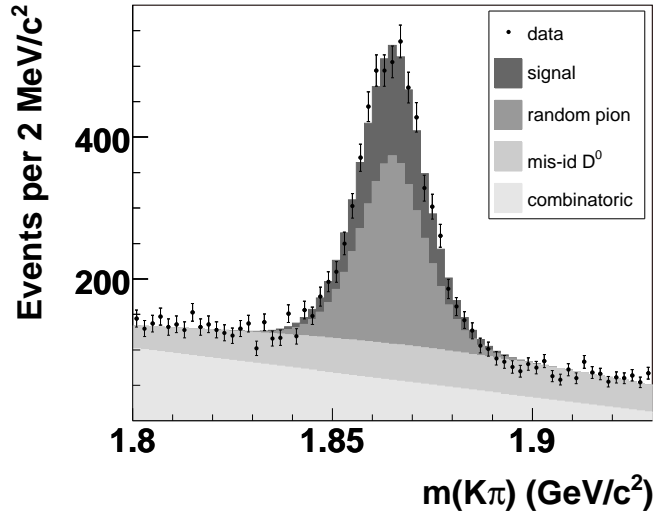}
\caption{$K\pi$ invariant mass distribution for candidates reconstructed
as $D^0 \to K^+\pi^-$ (DCS), requiring 
5~MeV/$c^2 < \Delta m < 7$~MeV/$c^2$. The shaded regions are projections
from the overall fit onto this distribution. This mass plot
illustrates the relative contributions
from the DCS signal and the three types of background, as described in
the text.}
\label{fig:proj}
\end{center}
\end{figure}

The $K\pi$ mass distribution for CF decays has been reported in a recent
CDF II publication on singly Cabibbo-suppressed decays \cite{cdf:kkpipi}. 
The $K\pi$ mass distribution for DCS candidates is illustrated by the histogram
in Fig.~\ref{fig:proj} for candidates satisfying 
5~MeV/$c^2 < \Delta m < 7$~MeV/$c^2$, where
$\Delta m = m(K^+\pi^-\pi^+) - m(K^+\pi^-) -m(\pi^{+})$.
Four categories of  $K\pi\pi$ combinations contribute to the distribution.
The first category (signal)
is DCS signal from $D^*$, with the correct $D^0 \to K^+ \pi^-$ 
interpretation. The second category (random pion) is background
from CF $D^0$ decays, where a randomly selected particle, 
usually from the primary interaction, is used as the tagging pion  
to form the $D^*$ candidate.  The third category (mis-id $D^0$) is background 
from $D^*$ decays where the $K$ and $\pi$ assignments 
from the CF $D^0$ decay are mistakenly interchanged. 
The last category is combinatoric background, where one or both tracks 
do not belong to a $D \to K\pi$ decay.

To determine the signal and background, candidates are divided into 60
slices of $\Delta m$, each slice of width 0.5~MeV/$c^2$.
(The distribution in Fig.~\ref{fig:proj} is for a 2~MeV/$c^2$
wide slice for purpose of illustration.) For each slice, the
$K\pi$ mass distribution is fit using a binned likelihood method with
a predetermined $D^0$ signal shape and a linear function for the
background.  The $D^0$ signal shape is determined from a fit to the CF
$K\pi$ distribution, which has a negligible background compared to the signal.
The fit range
(1.80 - 1.93)~GeV/$c^2$ excludes background from $D^0 \to K^+K^-$ and
$D^0 \to \pi^+\pi^-$ decays that are mis-reconstructed as $K^+\pi^-$.
The amplitude of the combined signal and random pion term and the
combinatoric background parameters are allowed to float independently
for each slice.  The amplitude and shape of the mis-id $D^0$
contribution is determined from the CF signal, by interchanging the pion
and kaon assignments.

\begin{figure}[!tbp]
\vskip +0.4cm
\begin{center}
\includegraphics[width=3.4in]{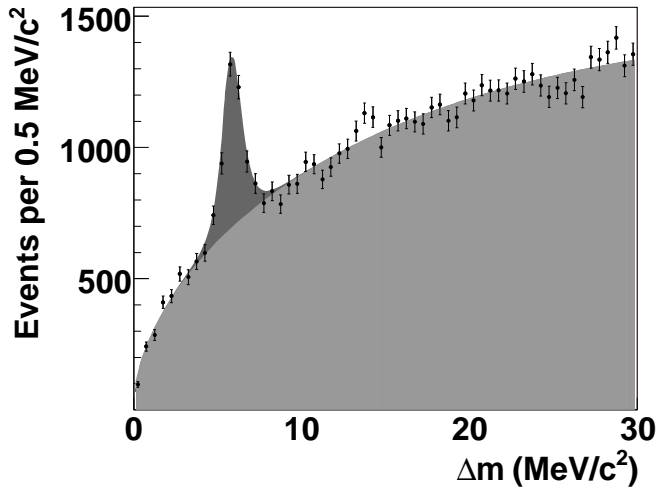}
\caption{
The number of $D^0 \to K^+\pi^-$ (DCS) decays as a function
of $\Delta m$.  The data points and statistical uncertainty bars 
are taken from the $K\pi$
slice fits. The shaded regions are determined from a least-squares fit
and show the contributions from signal (dark gray) and random tagging
pion background (light gray) as explained in the text.}
\label{fig:ws_yield}
\end{center}
\end{figure}

To determine the amount of DCS signal and random pion background, the
$D^0$ signal yields for the slices are plotted as a function of
$\Delta m$, as shown in Fig.~\ref{fig:ws_yield}. This distribution
is fit using a least-squares method with a signal shape predetermined
from the CF $\Delta m$ distribution and a background function of the
form $A (\Delta m)^B e^{-C (\Delta m)}$.  The amplitudes of the signal
and background terms and the background shape parameters $B$ and $C$
are determined from the fit. The fit results are $2005 \pm 104$ DCS
signal and $495172 \pm 907$ CF signal; their ratio gives
$ R_{\cal B} = (4.05 \pm {}$\mbox{0.21) $\times\, 10^{-3}$}.

\begin{table}[!!bp]
 \begin{center}
\caption{Dominant systematic uncertainties for $R_{\cal B}$. 
The sources lead to uncertainties in the DCS signal estimate.}
\label{tab:systematics}
  \begin{tabular}{ccccc}
   \hline \hline
      & Source   & &  Uncertainty ($\times\, 10^{-3}$) &    \\
     \hline
      & $K\pi$ combinatoric background shape  & &  0.09 &  \\
      & $\Delta m$ random pion background shape &  &  0.06 &  \\
      & $K\pi$ mis-ID $D^0$ background amplitude &  &  0.01 &  \\
     \hline \hline
  \end{tabular}
 \end{center}
\end{table}

Most of the detector properties that affect the DCS and CF signals
   are common and hence do not affect the ratio.  Thus, there are no
   systematic uncertainties due to geometric acceptance, particle
   identification, and trigger efficiency.
While the number of background events is
similar for the DCS and CF candidates, the size of the DCS signal is much
smaller. Thus, systematic uncertainty in the DCS background which
affects the DCS signal estimate also affects the ratio. There
are three such significant sources of systematic uncertainty, as
summarized in Table~\ref{tab:systematics}.  

To estimate the uncertainty due to the assumed combinatoric background
shape in the DCS $K\pi$ slice fits, we compared $R_{\cal B}$ results for two
shapes.  The nominal shape is linear and gives a good fit. We also
tried a quadratic form and assigned the change in $R_{\cal B}$ as a
systematic uncertainty. To estimate the uncertainty due to the assumed
$\Delta m$ background shape, we compared $R_{\cal B}$ results for two shapes.
The nominal shape is given by the function described earlier and gives
a good fit. We also tried a function with an additional parameter and
assigned the change in $R_{\cal B}$ as a systematic uncertainty. In fitting
the DCS $K\pi$ slice fits, the amplitude of the mis-id $D^0$
background is fixed from the CF signal. A simulation of the fitting
procedure is used to propagate the statistical uncertainty on the
background amplitude to a systematic uncertainty on $R_{\cal B}$.

We considered other sources of systematic uncertainty that we found to
be negligible. These include effects due to small differences in
detection efficiencies for $K^+$ versus $K^-$ and $\pi^+$ versus
$\pi^-$, which are reported in \cite{cdf:kkpipi}.
We tried alternative fits to the DCS $K\pi$ distributions
by extending the upper limit of the mass range from 1.80
to 2.00~GeV/$c^2$. This study required 
adding an explicit term for background from
$D^0 \to \pi^+\pi^-$ decays.

\begin{figure}[!t!p]
\vskip +0.4cm
\begin{center}
\includegraphics[width=3.4in]{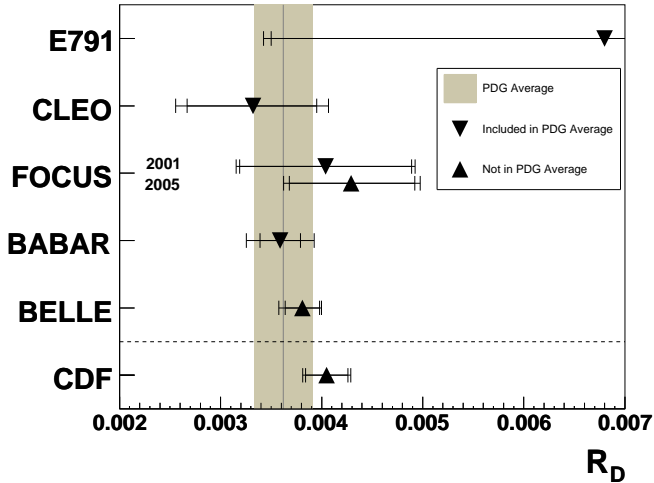}
\caption{Comparison of this measurement of $R_D$ with other recent results.
All the experimental fits assume no mixing or CP violation.  The inner
set of bars indicate statistical uncertainty; the outer set indicates the
quadratic sum of statistical and systematic uncertainties.  The shaded region
spans the PDG average and uncertainty~\cite{pdg04-mix}.  That average
includes measurements from E791~\cite{aitala:e791},
CLEO~\cite{godang:cleo}, FOCUS(2001)~\cite{link:focus}, and
BABAR~\cite{aubert:babar}.  The Belle~\cite{abe:belle},
FOCUS(2005)~\cite{link:focus05} and current CDF measurements are not
included in the PDG average.}
\label{fig:overview_comparison}
\end{center}
\end{figure}

In conclusion, we find $R_{\cal B} = [4.05 \pm 0.21 (\mbox{stat})
\pm {}$\mbox{0.11(syst)] $\times\, 10^{-3}$}. 
The difference between this value and the world average value for
$\tan^4\theta_C$ is
$(1.17 \pm {}$\mbox{0.34) $\times\, 10^{-3}$}, a 3.4 $\sigma$
deviation from zero.  If not a statistical fluctuation, this
difference could be due to violation of flavor SU(3) symmetry causing
$R_D \neq \tan^4\theta_C$, or could be a result of mixing.
While we cannot rule out the possibility of mixing from our result alone,
our result is consistent with the scenario of modest symmetry violation and
negligible mixing.  As shown in Fig.~\ref{fig:overview_comparison}, our
measured value of $R_{\cal B}$ is in fact consistent with the world
average and the most accurate individual measurements of $R_D$ obtained
from BABAR~\cite{aubert:babar} and Belle~\cite{abe:belle}.  Using the
technique we have established to extract the $D^0 \to K^+ \pi^-$ signal,
we can perform a time-dependent analysis using a larger data sample than
reported here, to separately measure $R_D$ and the mixing parameters
$x'$ and $y'$.

% comment out when using line numbers:
%\begin{acknowledgments}

We thank the Fermilab staff and the technical staffs of the
participating institutions for their vital contributions. This work
was supported by the U.S. Department of Energy and National Science
Foundation; the Italian Istituto Nazionale di Fisica Nucleare; the
Ministry of Education, Culture, Sports, Science and Technology of
Japan; the Natural Sciences and Engineering Research Council of
Canada; the National Science Council of the Republic of China; the
Swiss National Science Foundation; the A.P. Sloan Foundation; the
Bundesministerium f\"ur Bildung und Forschung, Germany; the Korean
Science and Engineering Foundation and the Korean Research Foundation;
the Particle Physics and Astronomy Research Council and the Royal
Society, UK; the Russian Foundation for Basic Research; the Comisi\'on
Interministerial de Ciencia y Tecnolog\'{\i}a, Spain; in part by the
European Community's Human Potential Programme under contract
HPRN-CT-2002-00292; and the Academy of Finland.

% comment out when using line numbers:
%\end{acknowledgments}


\begin{thebibliography}{99}

\bibitem{pdg04-mix}
S. Eidelman {\it et al.} (Particle Data Group), 
Phys.\ Lett.\ B {\bf 592}, 1 (2004),  
review by D. Asner on $D^0-\bar{D}^0$ mixing.

\bibitem{SU3-1} 
R. L. Kingsley, S. B. Trieman, F. Wilczek and A. Zee,
\prd{11}{1919}{1975}.

\bibitem{SU3-2} 
M. B. Voloshin, V. I. Zakharov and L. B. Okun,
\pisma{21}{403}{1975}{183}.

\bibitem{pdg04-vus} {\it Ibid.} [1], 
review by
E. Blucher and W.J. Marciano on $V_{ud}$, $V_{us}$, the Cabibbo angle, 
and CKM Unitarity.

\bibitem{Gronau-Rosner}
M. Gronau and J. L. Rosner,
Phys. Lett. B {\bf 500}, 247 (2001).

\bibitem{Golowich:2005pt}
E. Golowich and A. A. Petrov, 
%
%``Short distance analysis of D0 - anti-D0 mixing,'' 
%
Phys.\ Lett.\ B {\bf 625}, 53 (2005).
%
%[arXiv:hep-ph/0506185].

\bibitem{Burdman-Shipsey-2003}
G. Burdman and I. Shipsey, 
Annu. Rev. Nucl. Sci. {\bf 53}, 431 (2003). See section 2.3.

\bibitem{aubert:babar}
B. Aubert {\it et al.} (BABAR Collaboration), 
Phys.\ Rev.\ Lett. {\bf 91}, 171801 (2003).
%     [arXiv:hep-ex/0304007].

\bibitem{abe:belle}
K. Abe {\it et al.} (Belle Collaboration),
Phys.\ Rev.\ Lett. {\bf 94}, 071801 (2005).

\bibitem{CDF}
D.~Acosta {\it et al.} (CDF Collaboration),
Phys.\ Rev.\ D {\bf 71}, 032001 (2005).

\bibitem{cdf:silicon}
A. Sill {\it et al.},
Nucl.\ Instrum.\ Methods\ {\bf A447}, 1 (2000).

\bibitem{cdf:cot}
T. Affolder {\it et al.},
Nucl.\ Instrum.\ Methods\ {\bf A526}, 249 (2004).

\bibitem{cdf:xft}
E. J. Thomson {\it et al.}, 
IEEE Trans. Nucl. Sci. {\bf 49}, 1063 (2002).

\bibitem{cdf:svt} W. Ashmanskas {\it et al.}, Nucl.\ Instrum.\
Methods\ {\bf A518}, 532 (2004).

\bibitem{cdf:kkpipi} D. Acosta {\it et al.} (CDF Collaboration),
Phys.\ Rev.\ Lett.\ {\bf 94}, 122001 (2005).
% [arXiv:hep-ex/0504006].  

\bibitem{aitala:e791}
E.~M.~Aitala {\it et al.} (E791 Collaboration),
Phys.\ Rev.\ D {\bf 57}, 13 (1998).
%[arXiv:hep-ex/9608018].

\bibitem{godang:cleo}
R.~Godang {\it et al.} (CLEO Collaboration),
%``Search for $D^0 - \bar{D^0}$ Mixing",
Phys.\ Rev.\ Lett. {\bf 84}, 5038 (2000).
%[arXiv:hep-ex/0001060].

\bibitem{link:focus}
J.~M.~Link {\it et al.} (FOCUS Collaboration),
Phys.\ Rev.\ Lett. {\bf 86}, 2955 (2000).
%[arXiv:hep-ex/0012048].

\bibitem{link:focus05}
J.~M.~Link {\it et al.}  (FOCUS Collaboration),
% ``Measurement of the doubly Cabibbo suppressed decay D0 $\to$ K+ pi-
% and a search for charm mixing,''
Phys.\ Lett.\ B {\bf 618}, 23 (2005).
%[arXiv:hep-ex/0412034].

\end{thebibliography}
\end{document}